\documentclass[prl,twocolumn,floatfix,showkeys]{revtex4}

\usepackage{color}
\usepackage[normalem]{ulem}

\usepackage[english]{babel}
\usepackage{amssymb}
\usepackage{amstext}
\usepackage{graphicx}

\begin{document}

\title{Phenotypic switching can speed up biological evolution of microbes}

\author{Andrew C. Tadrowski$^1$}
\author{Martin R. Evans$^1$}
\author{Bart\l omiej Waclaw$^{1,2}$}
\affiliation{$^1$SUPA, School of Physics and Astronomy, The University of Edinburgh, Peter Guthrie Tait Road, Edinburgh EH9 3FD, United Kingdom\\
$^2$Centre for Synthetic and Systems Biology, The University of Edinburgh}

\begin{abstract}
Stochastic phenotype switching has been suggested to play a beneficial role in microbial populations by leading to the division of labour among cells, or ensuring that at least some of the population survives an unexpected change in environmental conditions. Here we use a computational model to investigate an alternative possible function of stochastic phenotype switching - as a way to adapt more quickly even in a static environment. We show that when a genetic mutation causes a population to become less fit, switching to an alternative phenotype with higher fitness (growth rate) may give the population enough time to develop compensatory mutations that increase the fitness again. The possibility of switching phenotypes can reduce the time to adaptation by orders of magnitude if the ``fitness valley'' caused by the deleterious mutation is deep enough. Our work has important implications for the emergence of antibiotic-resistant bacteria. In line with recent experimental findings we hypothesise that switching to a slower growing but less sensitive phenotype helps bacteria to develop resistance by exploring a larger set of beneficial mutations while avoiding deleterious ones.
\end{abstract}
\keywords{evolution $\vert$ population genetics $\vert$ stochastic phenotype switching}
\maketitle

\section{Introduction}
Biological evolution relies on two mechanisms which are instrumental in natural selection: preferential survival of better adapted individuals (selection) and variations among individuals (phenotypic variation). One of the sources of phenotypic variability is genetic alteration due to mutations and recombination. However, even genetically identical organisms will often behave differently because the same genotype may lead to many different phenotypes - observable traits of an organism - as a result of environmental factors and the organism's history. Although ubiquitous and easily observed in animals and plants, phenotypic diversity can already be amply demonstrated  in microorganisms. Examples range from different cell sizes depending on growth medium \cite{yao_regulation_2012}, through bistability in utilization of different food sources \cite{ozbudak_multistability_2004}, to diversification between motile/non-motile cells \cite{kearns_cell_2005}. Microorganisms are often able to switch between these phenotypes in response to a change in external conditions such as the arrival of a new food source or depletion of the currently used one. A typical example is diauxic shift - a switch to another food source, for example from glucose to cellobiose in {\it L. lactis} when glucose becomes depleted \cite{solopova_bet-hedging_2014}, which involves altering gene expression levels without changing the genetic code.

\begin{figure}
   \centering
    \includegraphics[width=0.45\textwidth]{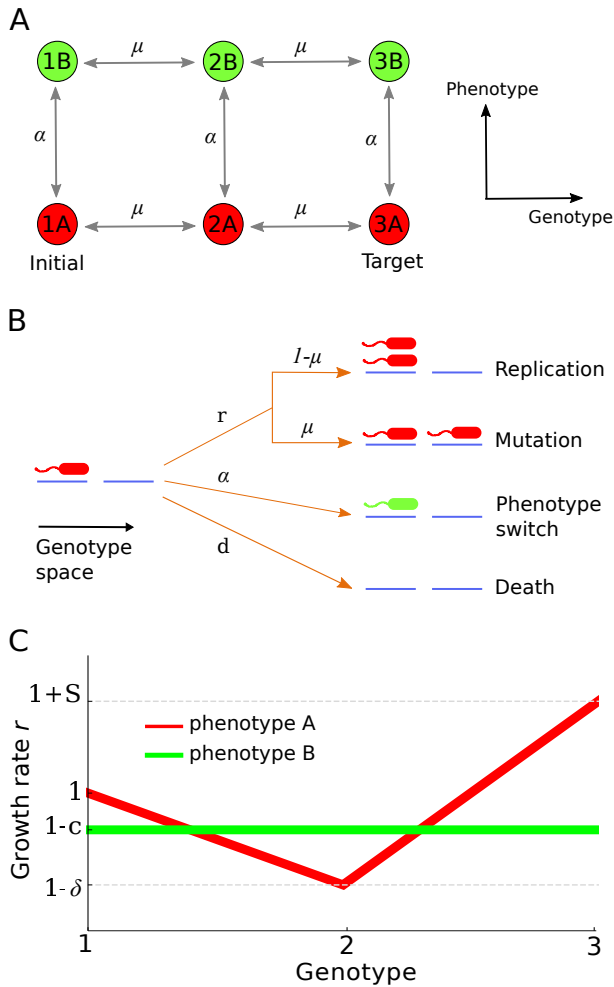}\\ 
    \caption{\label{fig:model}The model. (A) Diagram showing the six possible states of a cell and the available transitions between them. The genotypes are labelled 1, 2 and 3, the phenotypic states are labelled A and B. Transitions between genotypes/phenotypes occur at rates $\mu$ and $\alpha$, respectively. All cells are initially in state 1A. Evolution continues until a single cell reaches the target state 3A. (B) Each cell can replicate, switch phenotype, or die with rates $b,\alpha$ and $d$ respectively. Upon replication a cell has the probability $\mu$ of producing a mutant of each neighbouring genotype. (C) The fitness landscapes for both phenotypes. Phenotype A has a fitness valley at 2A while phenotype B has uniform fitness across all genotypes.}
\end{figure}

Some microorganisms switch seemingly randomly between two or more phenotypes even in the absence of external stimuli. This causes the population to become phenotypically heterogeneous. Several explanations have been proposed as to why \textit{stochastic phenotype switching} has evolved \cite{ackermann_functional_2015}. One of them is the division of labour \cite{ackermann_self-destructive_2008} in which different microbial cells perform different functions and this maximizes the benefit to the population. Another theory, called bet hedging \cite{kussell_phenotypic_2005,beaumont_experimental_2009}, proposes that in a fluctuating and unpredictable environment it pays to have a fraction of the population in a different state, which is perhaps maladapted to the present environment but better suited to possible future environments. Since only a small fraction of the population expresses the maladapted phenotype at any one time,
this strategy conserves resources while allowing the population to stay prepared for an unexpected change. Examples include bacterial persisters \cite{balaban_bacterial_2004, kussell_bacterial_2005}, flu(Ag43)/fim switch \cite{hasman_antigen_2000,van_der_woude_regulation_2008,balaban_bacterial_2004,adiciptaningrum_direct_2009} and competence to non-competence switching in the bacterium \textit{B. subtilis} \cite{maamar_bistability_2005}.
Animal cells also exhibit this behaviour. Epithelial-to-mesenhymal transition -- a landmark in cancer progression -- is thought to be a phenotypic, epigenetic change \cite{mcdonald_genome-scale_2011}. There is also some evidence that phenotypic switching may be involved in resistance to chemotherapy \cite{kemper_phenotype_2014}, although this remains controversial \cite{zapperi_cancer_2012}.

Here we investigate an alternative possible role for the evolution of phenotype switching in microbes: the facilitation of genetic evolution in a static environment. Previous theoretical work \cite{levin_non-inherited_2006} has demonstrated that switching to a persistent phenotype can provide a larger pool of bacteria to evolve genetic resistance to antibiotics. That work only addressed the situation in which resistance is caused by a single, beneficial mutation. However, a significant change of fitness, defined here as a measure of a microorganism's growth rate, often requires multiple mutations. Although many fitness landscapes have at least one accessible pathway \cite{weinreich_darwinian_2006, franke_evolutionary_2011}, along which the fitness increases monotonically, in some cases a fitness valley - a genotype with a lower fitness than all its neighbours - is unavoidable. For example, developing resistance to the antibiotic streptomycin involves a fitness cost which must be counteracted by compensatory mutations, i.e.  subsequent mutations that increase the microorganism's fitness to at least that of the ancestral strain \cite{schrag_reducing_1996, schrag_adaptation_1997, maisnier-patin_compensatory_2002}.

Here we study evolution in the presence of such a fitness valley, in which a population must acquire two mutations to reach the best-adapted genotype. The first mutation is deleterious and corresponds to a valley in the fitness landscape that is crossed by a second, compensatory mutation. Theory and computer modelling suggests that fitness valleys significantly affect biological evolution \cite{weissman_rate_2009, weinreich_rapid_2005, covert_experiments_2013, szendro_predictability_2013}, often, but not always, slowing it down. However, in our model cells can also switch to an alternative phenotype in which the fitness landscape is flat. We show that this switching enables the population to avoid the costly valley, even if the alternative phenotype is less fit than the initial state. We also demonstrate the existence of an optimal range for the switching rate. In this range the time to evolve the best-adapted genotype can be reduced by many orders of magnitude compared to the case without switching. Finally, we show that if the switching rate is allowed to evolve it will converge to values within this optimal range.

\section{Model}
We consider a population of haploid cells. Each cell can be in one of two phenotypic states, A and B, and can assume one of three genotypes (``genetic states''), as shown in Fig.~\ref{fig:model}A,B. Cells replicate stochastically with rate $r_i(1-N/K)$, where $i$ labels one of the six possible states, $N$ is the total population size and $K$ is the carrying capacity of the environment. The logistic-like factor $(1-N/K)$ causes growth to cease when $N$ becomes as large as $K$, which limits the total population size. During replication a cell will produce a mutated offspring of an adjacent genotype with probability $\mu$. Mutation does not change the phenotypic state. All cells switch randomly between the states A and B at the symmetric switching rate $\alpha$. Cells are randomly removed from the population at a constant death rate $d$. If the carrying capacity $K$ is small ($K = 10\cdots 10^4$) the model is appropriate to describe a small microbial population growing in a microfluidic chemostat with constant dilution rate \cite{balagadde_long-term_2005}. For larger $K$ ($K = 10^4 \cdots 10^9$) the model is relevant to populations cultured in mesoscopic (cm-size) chemostats.

The population initially consists of all individuals of type 1A, i.e. of genotype $1$ in phenotypic state A, which has the growth rate $r_{\rm 1A}=1$ (arbitrary units). The initial population size, unless otherwise stated, is equal to the equilibrium size $(1-d)K$. State 3A is the global maximum of both fitness landscapes with the growth rate $r_{\rm 3A}=1+S$, where $S>0$ is its selective advantage over the `wild-type' 1A (Fig.~\ref{fig:model}C). Such a beneficial mutant has the probability of fixation, from a single individual in a population of cells in state 1A, approximately given by   $\frac{1-e^{-S}}{1-e^{-K S}}\approx S$ \cite{nowak_evolutionary_2006}  when $1/K\ll S\ll 1$. We are interested in the {\it adaptation time} $T$ that it takes for the population to evolve the first individual in state 3A, conditioned on the population not going extinct (otherwise the time would be infinite).

We shall begin by considering the case in which the growth rate $r_{\rm 2A}=1-\delta$, while the growth rates $r_{\rm 1B}=r_{\rm 2B}=r_{\rm 3B}=1$, which leads to a ``fitness valley'' in phenotype A and a flat fitness landscape in phenotype B (Fig.~\ref{fig:model}C). In the absence of phenotype switching the only way a population in state 1A can evolve an individual in state 3A is to go through the low-fitness state 2A. However, phenotype B allows the population to traverse an alternative route that avoids state 2A.

\section{Results}
\subsection{Optimal range of the switching rate exists in which evolution is fastest}

\begin{figure}
   \centering
    \includegraphics[width=0.45\textwidth]{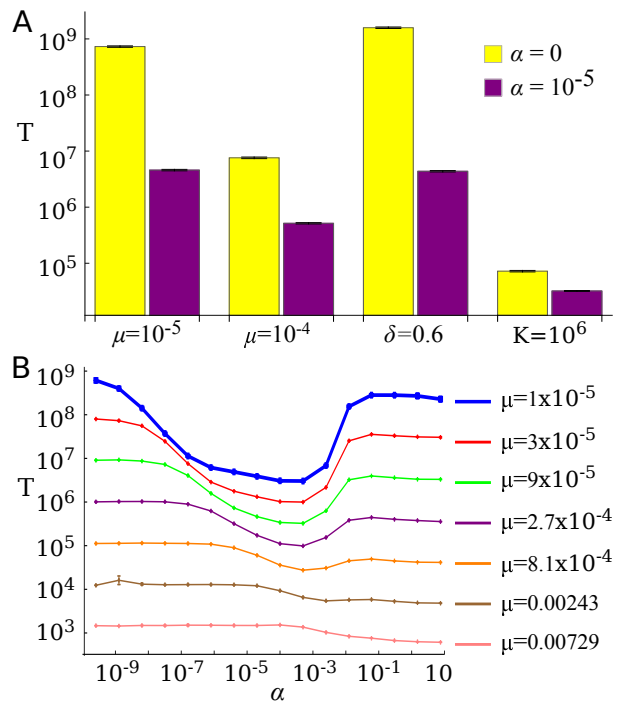}
    \caption{\label{fig:results}Adaptation time $T$ for different scenarios -- note logarithmic scale. (A) A bar chart comparing pairs of $T$ values with and without switching phenotypes ($\alpha=10^{-5}$ and $\alpha=0$ respectively) for different parameter values. Left-most pair of bars: $K=100, \mu=10^{-5}, \delta=0.4$ and $d=0.1$. A label underneath each pair of bars indicates which variable has been changed compared to the left-most pair. (B) $T$ as a function of the switching rate $\alpha$ for a range of mutation probabilities $\mu$. Parameters as in (A). For small enough $\mu$ an optimal (minimizing $T$) switching rate can be seen.}
\end{figure}

Figure~\ref{fig:results}A shows simulation results examining the effect of phenotype switching on the adaptation time $T$ for a range of the parameters of the model. In all cases the presence of low frequency switching, i.e. small $\alpha$, decreases the adaptation time by orders of magnitude. Holding all other parameters fixed and changing only the switching rate reveals an optimal range for $\alpha$ (Fig.~\ref{fig:results}B) in which $T$ is minimized provided $\mu$ is sufficiently small. For the smallest mutation probability ($\mu=10^{-5}$) considered in Fig.~\ref{fig:results}B the minimal adaptation time is approximately two orders of magnitude lower than in the absence of switching ($\alpha=0$), or when switching is frequent ($\alpha$ very large). Only if the mutation rate is unrealistically large does switching not speed up evolution (SI and Fig. S1).

In the absence of transitions between states 2A and 2B (i.e. removing this step from the model in Fig. \ref{fig:model}A) the adaptation time decreases monotonically with $\alpha$ (SI and Fig. S2).

\subsection{Fastest evolutionary trajectories avoid the valley}
\begin{figure*}
   \centering
    \includegraphics[width=1.0\textwidth]{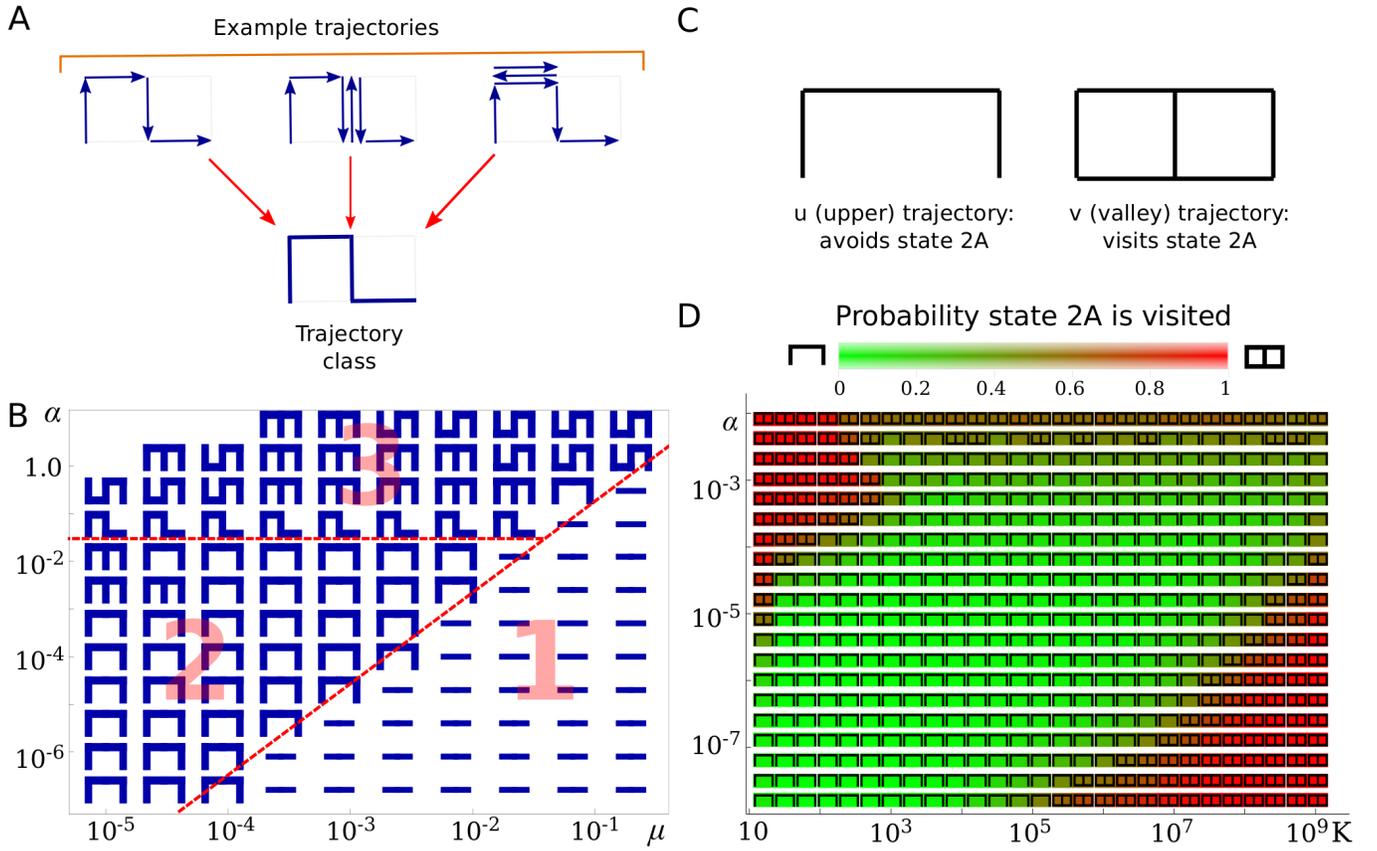}\\
    \caption{\label{fig:trajectory}Trajectories of successful cells in genotype/phenotype space. (A) Diagram showing how different trajectories that go through the same set of states are grouped together into the same trajectory class. This class is represented as a symbol in which blue lines correspond to transitions made by the successful cell. (B) The most probable trajectory class as a function of $\mu$ and $\alpha$, for $K=100$. Three regions labelled $1,2,3$ can be distinguished. (C) Trajectories can be further grouped regarding whether they avoid (trajectory ``u'') or visit (trajectory ``v'') the fitness valley at state 2A. (D) The most common trajectory group (symbols as in panel C) and the probability of a trajectory visiting state 2A (colours, see the colour bar) as a function of switching rate $\alpha$ and carrying capacity $K$, for $\mu=10^{-6}$. In all simulations $\delta=0.4$ and $d=0.1$.}
\end{figure*}

To understand the evolutionary trajectory selected in the optimal range of switching frequency we examined the histories of successful cells, i.e. the states from Fig. \ref{fig:model}A visited during evolution from state 1A to the final state 3A. We then represented each cell's trajectory in the state space as a sequence of links connecting the visited states (Fig. \ref{fig:trajectory}A). These were then classified as one of 21 classes by grouping sequences with multiple (back-and-forth) transitions between the states together with those without (Fig. \ref{fig:trajectory}A). Fig. \ref{fig:trajectory}B shows the most probable trajectory as a function of $\mu$ and $\alpha$. As expected, simulated evolution favours different trajectories depending on the values of $\mu,\alpha$. The $(\mu,\alpha)$-space can be approximately separated into three regions, each corresponding to one or more trajectory classes. 
Region $1$  corresponds to trajectories that go through the fitness valley at 2A. In this region mutations are frequent enough to offset the loss of fitness incurred when passing through the valley. In region $2$ the dominant trajectory avoids the valley by switching to phenotype B. This region corresponds to the fastest adaptation times from Fig. \ref{fig:results}B. Region $3$ is characterised by a mixture of trajectory types using both phenotype B and visiting state 2A. This region corresponds to large $\alpha$ which not only enables transitions to phenotype B but also from state 2B to the deleterious state 2A. The latter is responsible for the increase in the adaptation time for large $\alpha$ in Fig. \ref{fig:results}B. A similar decomposition into three regions can be obtained from the probability of individual transitions occurring between the states of the system (Fig. S3).

We next examined which trajectories would be preferred if we varied $K$ instead of $\mu$, as the carrying capacity is more likely to vary significantly in real microbial populations and is much easier to control experimentally than the mutation rate. Since we ascertained earlier (cf. the previous paragraph) that the time to adaptation is most significantly affected by whether the trajectory visits state 2A or not, we further reduced the number of trajectory classes to just two (Fig. \ref{fig:trajectory}C): the ``u'' (upper) trajectory type that avoids state 2A or the ``v'' (valley) trajectory type that visits it. We found a large region in the space $(K,\alpha)$ that favours trajectories that avoid the state 2A (Fig. \ref{fig:trajectory}D). This region extends to large values of $K\sim 10^{9}$. The adaptation time depends non-monotonically on the population size $K$(Fig. S4); similar non-monotonic behaviour has been seen in models without phenotype switching \cite{martin_population_2007,elena_effects_2007}.

\subsection{Results are robust to small fitness costs of phenotype switching}

\begin{figure}
   \centering
    \includegraphics[width=0.48\textwidth]{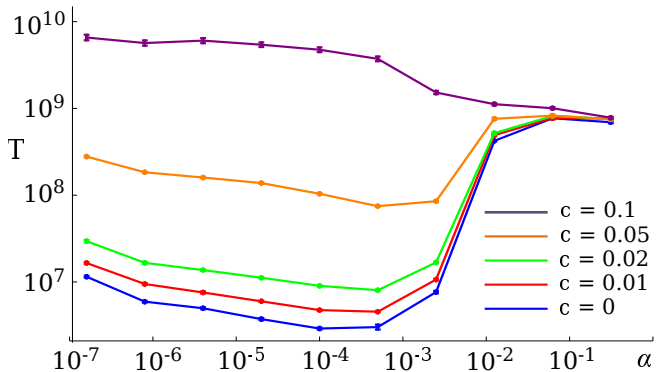}\\    
    \caption{\label{fig:cost}Adaptation time in the presence of fitness costs. The adaptation time $T$ as a function of the switching rate $\alpha$ for different fitness costs $c=\{0,0.01,0.02,0.05,0.1\}$ of phenotype B and for parameters $K=100, \mu=10^{-5},\delta=0.9$ and $d=0.1$.}
\end{figure}

So far we have assumed that the alternative phenotype B has the same fitness as the wild-type state 1A. We shall now consider the case in which phenotype B is less adapted to the environment than the state 1A. This is a common scenario; for example, phenotypes more resistant to antibiotics often have a lower growth rate than susceptible phenotypes do in the absence of the drug \cite{balaban_bacterial_2004}. To model this, we assume that the growth rates $r_{\rm 1B}=r_{\rm 2B}=r_{\rm 3B}=1-c$, where $c>0$ is the fitness cost of switching to phenotype B (see Fig. \ref{fig:model}C). A cell that switches from 1A to 1B grows more slowly but switching can still be advantageous if the alternative route by phenotype B is faster than crossing the fitness valley at state 2A. Fig. \ref{fig:cost} shows the adaptation time $T$ as a function of switching rate $\alpha$ for different fitness costs. For sufficiently small $c$ the same qualitative behaviour as for $c=0$ is observed.

The same optimal switching rate range as in Fig. \ref{fig:results} can be seen in Fig. \ref{fig:cost} for costs $c\leq 0.05$ of state 1A's fitness, although it reduces in significance as $c$ increases. This is due to the increased time taken by trajectories that avoid state 2A as a result of the reduced growth rates of the phenotype B states. For large $c$, such as $c=0.1$ in Fig. \ref{fig:cost}, there is no optimal switching range but the adaptation time $T$ decreases monotonically with $\alpha$. Stochastic phenotype switching thus speeds up evolution even in the presence of (moderate) fitness costs.

\subsection{Switching rate evolves until it reaches the optimal range}
So far we have shown that there often exists an optimal range for the switching rate that results in the shortest adaptation time. This range coincides with successful cells avoiding the deleterious state 2A by switching to phenotype B. To 
see whether a variable switching rate would evolve to be in this optimal range, thus optimising the evolutionary process, we simulated a population of cells that started at 1A with $\alpha\approx 0$ initially. The switching rate was subsequently allowed to evolve: during replication, offspring were assigned a new $\alpha$ value with probability $\mu$, the same as for mutations between genotypes $1\leftrightarrow 2\leftrightarrow 3$. The new $\alpha$ was randomly and uniformly selected from a fixed set of possible values ($\alpha=2.56\times10^{-10}\times5^{i}$ for integer $i \in [0,15]$); similar results were obtained when $\alpha$ evolved incrementally (Fig. S5). In this extended model the evolutionary trajectory of successful cells spans three dimensions: two for the genotype, one of which evolves the switching rate, and one for the two phenotypes A and B (Fig. 5A). The end point of the trajectory remains a cell produced at state 3A regardless of its switching rate. We collected trajectories of successful cells in the expanded state space, including states with different $\alpha$, and calculated transition probabilities between any two states. Fig. 5B shows these probabilities as links of different thickness between the states in state space. We observe that successful trajectories are unlikely to feature genotype mutations in phenotype A (very few red links). Instead, most successful cells first switch to state 1B, mutate to 3B, and switch back to state 3A. 
The evolved switching rate often falls within the optimal range found in Fig. \ref{fig:results}B for the same set of parameters $K,\mu$ and $d$. This is further illustrated in Fig. 5C where we overlay the adaptation time from Fig. \ref{fig:results}B with a bar chart that shows the probability that a particular $\alpha$ was selected by evolution.

Fig. 5C reveals two more interesting features. First, we see that cells that have evolved $\alpha$ above the optimal range ($\alpha \geq 0.02$) are unlikely to be successful due to fast transitions between states 2B and 2A (fitness valley). Such transitions effectively lower the fitness of state 2B and create a new fitness valley at this state. Second, we see that a large number of trajectories cross genotype space at $\alpha$ values below the optimal $\alpha$ range. However, the plot of $T$ in Fig. 5C suggests this will be a slow process. The latter observation  can be explained through these cells first evolving a large switching rate, allowing them to switch quickly to phenotype B, before their switching rate decreases again. This picture is corroborated by observing that at low $\alpha$ there is little phenotype switching (purple links in Fig. 5B) and yet the likelihood of mutations in phenotype B across genotype space is large (green  links in Fig. 5B). This provides an alternative efficient trajectory for a population to cross the fitness valley. However, such trajectories need at least one extra mutation compared to those that evolve $\alpha$ directly to within the optimal range. Therefore evolution is dominated by the latter trajectories, in particular for low mutation probabilities $\mu$. 

\begin{figure*}
   \centering
   \includegraphics[width=1.0\textwidth]{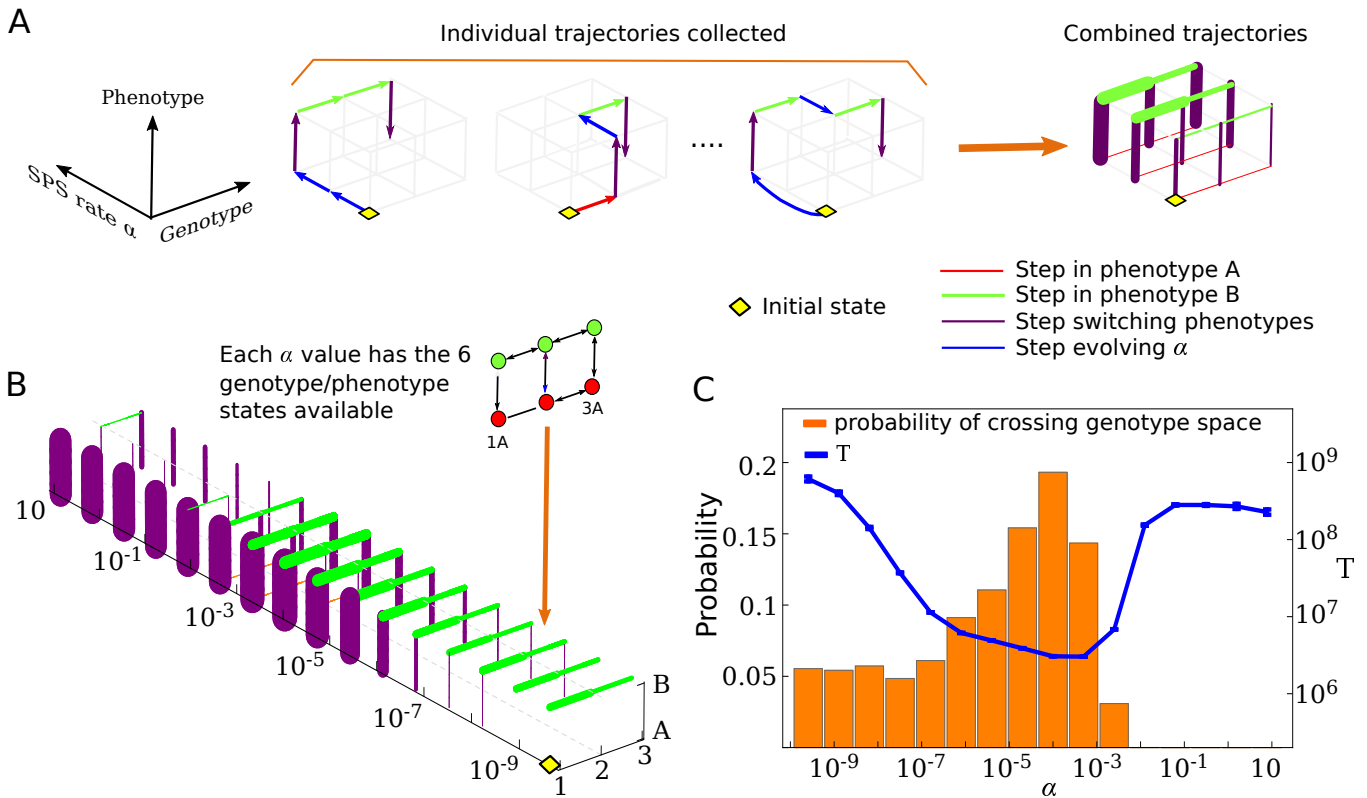}    
    \caption{Evolution selects switching rates within the optimal range. (A) Left: examples of evolutionary trajectories in the extended model in which the stochastic phenotype switching (SPS) rate $\alpha$ also evolves. (A) Right: trajectories are used to calculate transition probabilities between the states of the system, which are then represented by the thickness of links connecting the states. Red links correspond to mutations in phenotype A, green links to mutations phenotype B and purple links to switching between phenotypes. (B) Graph of transition probabilities where line thicknesses are proportional to the probability that a successful trajectory will take that step. The parameters are $K=100, \mu=10^{-5}, \delta=0.4$ and $d=0.1$, the same that were used in Fig. \ref{fig:results}B (blue line). The population begins at the wild-type 1A with $\alpha=2.56\times10^{-10}$ and evolves until a cell in state 3A is produced. (C) The probability that genotype space is crossed at a given $\alpha$ in either phenotype A or B. The probability has a maximum where the adaptation time $T$ for fixed $\alpha$ (plot from Fig. \ref{fig:results}B superimposed on the same graph) has its minimum.}
   \label{fig:EvolvingAlpha}
\end{figure*}

\section{Discussion}
Phenotypic plasticity has been recently shown to affect Darwinian evolution of animals \cite{Ghalambor_2015}.
Our theoretical research suggests that stochastic phenotypic switching is also able to significantly speed up biological evolution of microbes, perhaps by many orders of magnitude. Switching phenotypes enables cells to evade a fitness valley which may be difficult to cross otherwise, particularly for small populations. Although our model uses only three genotypes, we believe the results presented here are much more general and apply to realistic, multi-dimensional fitness landscapes \cite{szendro_quantitative_2013,palmer_delayed_2015,weinreich_darwinian_2006}. Epistasis causes such landscapes to contain many local fitness maxima separated by genotypes with lower growth rate \cite{Korona_1994}. In a small population, in the absence of phenotype switching, evolution can spend much time in a local maximum before ``tunnelling'' through one of the fitness valleys. Phenotypic switching opens up a second fitness landscape in which minima and maxima can be located at different genotypes compared to the first landscape.  Phenotype switching provides alternative routes between otherwise isolated fitness maxima in both landscapes, and thus enables evolution to proceed faster. This mechanism, similar to the proposed interplay between genetic and epigenetic mutations \cite{Klironomos_2013}, adds another dimension to the complexity of evolutionary pathways \cite{poelwijk_evolutionary_2006,weinreich_darwinian_2006,palmer_delayed_2015} which is relevant for the predictability of evolution \cite{de_visser_empirical_2014}.

We now discuss how plausible our computational results are for real microbial populations. Although our theory is quite general, we focus on antibiotic resistance evolution. We believe our model represents best a situation in which a population of microbes is exposed to sub-lethal concentrations of an antibiotic. In this scenario, state 1A corresponds to a `wild-type' genotype whose growth rate is slightly lowered by the presence of the antibiotic, but cells are still able to reproduce. The fitness landscape (Fig. \ref{fig:model}C) for phenotype A then corresponds to a situation in which an initial mutation further lowers the growth rate by e.g. making the target enzyme less susceptible to the drug but also less efficient \cite{schrag_adaptation_1997,schrag_reducing_1996,Marcusson_2009}, while a second mutation compensates for the loss of efficiency and increases the growth rate beyond that of the ancestral strain. In the same example the flat landscape for phenotype B can result from switching to a state with higher gene expression of multidrug efflux pumps or antibiotic-degrading enzymes. This has been observed for a biofilm-forming bacterium {\it P. aeruginosa} which can switch to a resistant phenotype in the presence of antibiotics \cite{breidenstein_pseudomonas_2011, drenkard_antimicrobial_2003}. Alternatively, phenotype B can be a persister - although the conventional view is that persistent cells do not replicate, this has been recently challenged \cite{wakamoto_dynamic_2013}. Slower (but non-zero) replication rates can enable persisters to mutate and explore a larger set of genotypes than the non-persister (phenotype A) would be able to do in the absence of switching.

In microbes, the probability of single nucleotide mutation is $10^{-10}-10^{-9}$ per replication \cite{drake_constant_1991,lee_rate_2012} but can be higher ($10^{-7}$) in mutator strains \cite{lee_rate_2012,sniegowski_evolution_1997} or in the presence of antibiotics \cite{gillespie_effect_2005}. Assuming $\mu=10^{-6}$ which would correspond to $\sim 10$ point mutations in a mutator strain or any loss-of-function mutation in a non-mutator strain (assuming a typical gene length 1kbp \cite{kerner_proteome-wide_2005}), Fig. \ref{fig:trajectory}D indicates that phenotype switching would be the preferred mode of evolution for $\alpha<10^{-4}$ in a population of $K=10^3$ cells, for $10^{-7}<\alpha<5\times 10^{-3}$ for intermediate size $K=10^6$, and for $\alpha\approx 10^{-3}$ for $K=10^9$. For lower mutation probabilities $\mu$ we expect this behaviour to extend to lower switching rates, consistent with the increasing optimal ranges seen in Fig. \ref{fig:results}B, while maintaining approximately the same upper limit of $\alpha \approx 10^{-3},\dots,10^{-2}$. Switching rates encountered in nature can be as large as this \cite{van_der_woude_regulation_2008,adiciptaningrum_direct_2009} and thus we believe that stochastic phenotypic switching can be a significant feature affecting the biological evolution of microbes.

The model we have proposed is a mathematical idealisation. Although it may be possible to test the quantitative predictions of our model in a carefully designed experiment, many of the assumptions we made would have to be relaxed to model naturally occurring populations of bacteria. However, recent work suggests that mechanisms similar to what we described here may be at play in real microbial populations. In particular, Refs. \cite{bos_emergence_2015,zhang_acceleration_2011} show that the bacterium {\it E. coli} switches to a filamentous phenotype (``phenotype B'') as a result of exposure to sub-inhibitory concentration of the antibiotic ciprofloxacin, and filamentous cells occasionally produce normal-length cells (``phenotype A'') resistant to ciprofloxacin. It has been hypothesized that phenotype B provides a ``safe niche'' giving bacteria enough time to evolve resistance. Although the switching between the phenotypes in this experiment is not entirely random but a response to stress, the observed behaviour is similar to that predicted by our model.

\section{Methods}

Custom written Java programs were used to simulate the model. For smaller $K$ ($K \leq 100$) each cell was assigned a current type (one of the 6 possible states from Fig. \ref{fig:model}A) and a sequence of past states beginning at state 1A. A variant of the Gillespie kinetic Monte Carlo algorithm\cite{gillespie_exact_1977} was used to evolve the system. In each time step, a random cell and action (replication, death, or phenotype switch) was selected with a probability proportional to the rate with which that action occurs. A second random number drawn from an exponential distribution, with mean equal to the inverse of the total rate of all possible actions in the system, was used as a measure of time elapsed during that step. For larger $K>100$ (or when $c>0$) the evolution of the system was approximately modelled using a type of $\tau-$leaping algorithm \cite{gillespie_approximate_2001}.

Statistical analysis was performed in Mathematica using a custom-written code. Mean values presented in all plots were obtained by averaging over $10^4$ runs, except for Fig. \ref{fig:trajectory}D where points were averaged over $10^{2}-10^{3}$ runs. Error bars are s.e.m.

\section*{Acknowledgments}
We thank Rosalind Allen for helpful discussions and many suggestions at the beginning of this project. BW was supported by a RSE/Scottish Government Personal Research Fellowship. MRE and AT thank the EPSRC for support through  grant number EP/J007404/1 and a studentship respectively. This work has made use of the resources provided by the Edinburgh Compute and Data Facility (ECDF) (http://www.ecdf.ed.ac.uk/).

\bibliographystyle{plain}
\bibliography{refs_new}{}

\section{Supplementary Information}

\setcounter{figure}{0}
\renewcommand{\thefigure}{S\arabic{figure}}

\subsection{Phenotypic switching will not always speed up evolution}

\begin{figure}[h]
   \centering
    \includegraphics[width=0.48\textwidth]{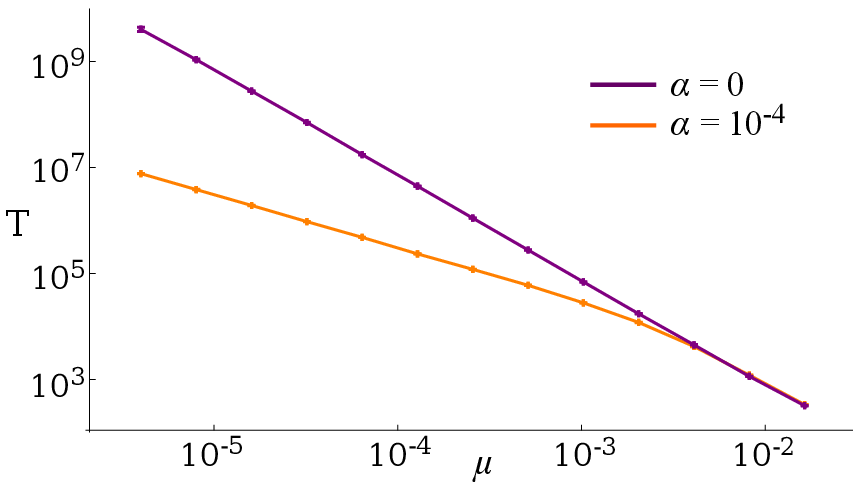}
    \caption{Phenotypic switching does not speed up evolution for large mutation probabilities $\mu$. Shown is the adaptation time $T$ from computer simulations as a function of $\mu$ with and without phenotype switching ($\alpha=10^{-4}$ and $\alpha=0$ respectively), for $K=100, \delta=0.4$ and $d=0.1$.}
    \label{fig:switching}
\end{figure}

In Fig. 2A we compared the adaptation time $T$ with and without phenotypic switching for various parameters. In all cases switching decreased $T$. However, if the mutation probability $\mu$ is large enough, the pathway that goes directly through the valley genotype 2A may take less time than any pathway involving the alternative phenotype B. Figure \ref{fig:switching} shows $T$ as a function of mutation probability $\mu$ for the case with and without switching. For large enough $\mu$, switching makes no observable difference to the adaptation time as successful cells do not use an alternative pathway. This is the same behaviour observed in region 1 of Fig. 3B.

\subsection{Adaptation time decreases monotonously with $\alpha$ in the absence of switching between states 2A and 2B}

\begin{figure}[h!]
   \centering
   \includegraphics[width=0.48\textwidth]{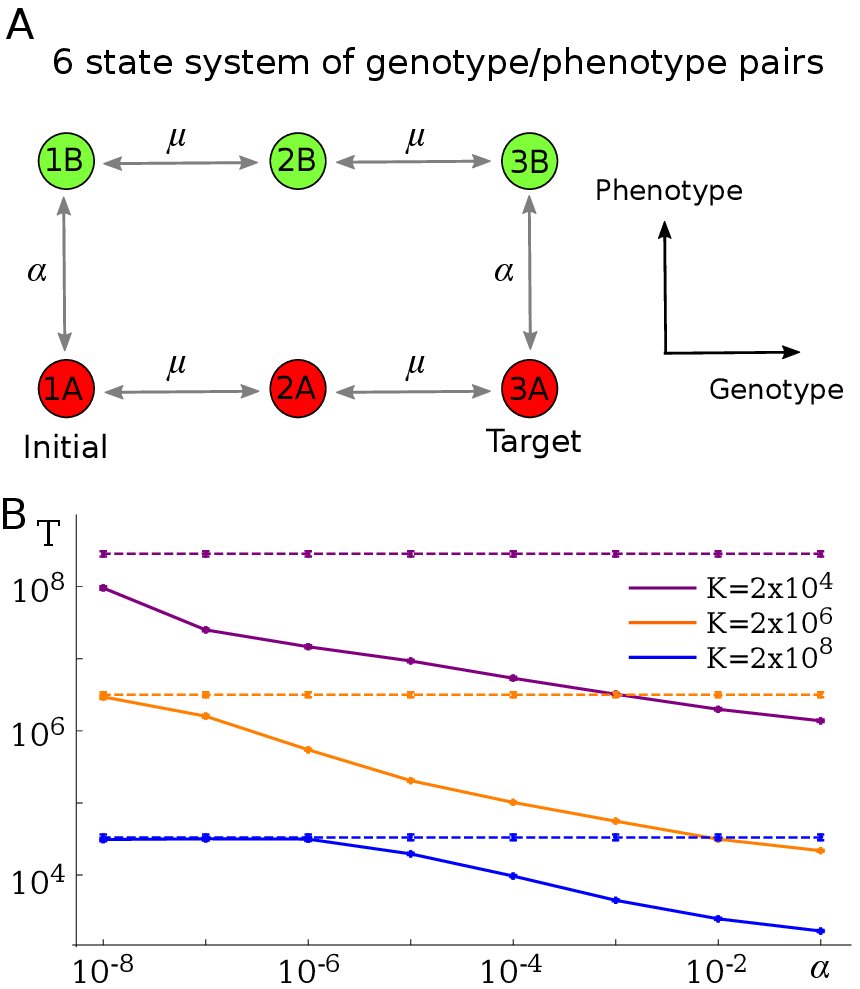}
    \caption{\label{fig:systemB}Time to adaptation without phenotype switching between states 2A and 2B. (A) The modified model. (B) Adaptation time $T$ as a function of switching rate $\alpha$ for $K=2\times10^{4}, 2\times10^{6}$ and $2\times10^{8}$. Dashed lines indicate $T$ when $\alpha=0$. Remaining parameters are $\mu=10^{-6}, \delta=0.4$ and $d=0.1$}
\end{figure}

We identified that region 3 in Fig. 3B appears due to the existence of the step between states 2A and 2B. For large $\alpha$ values a trajectory via phenotype B is likely to involve switching to the deleterious state 2A. This reduces the average fitness of the population and causes the adaptation time to increase with $\alpha$ (Fig. 2B). To demonstrate this consider the model without switching between states 2A and 2B (Fig. \ref{fig:systemB}A). The adaptation time $T$ as a function of switching rate $\alpha$ decreases monotonically in this modified model (Fig. \ref{fig:systemB}B). 

\subsection{Probability of individual steps for successful trajectories}
\begin{figure}
   \centering
    \includegraphics[width=0.48\textwidth]{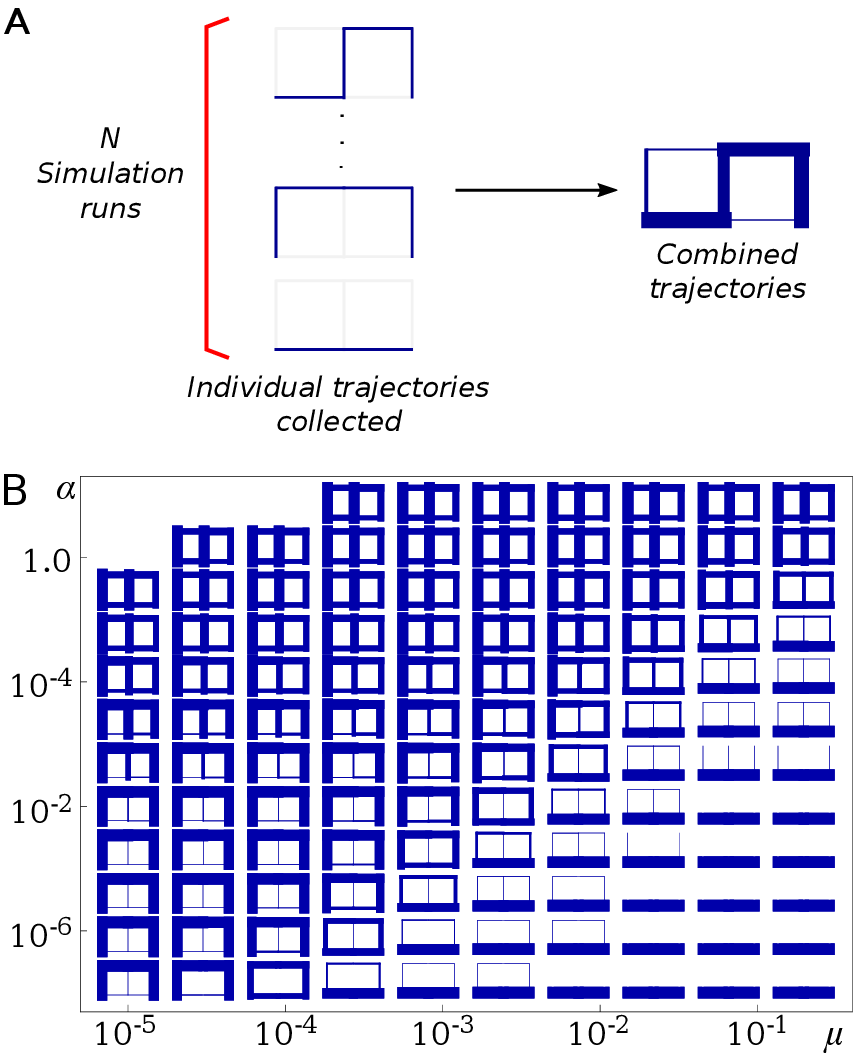}
    \caption{The probability of each step being taken by successful cells. (A) Constructing images that display the probability that each step is taken. By collecting many successful trajectory classes we can compile them into a single image that consists of links of different thicknesses, where the thickness of each link is proportional to the probability that that step is taken by a successful cell. (B) The probability each step is taken as a function of $\mu$ and $\alpha$. Remaining parameters are $K=100, \delta=0.4$ and $d=0.1$.}
    \label{fig:stepProbability}
\end{figure}

In Fig. 3B we showed the most common trajectory classes for successful cells as a function of $(\mu,\alpha)$. A more detailed picture can be obtained by looking at the probability that each step occurs in a successful cell's trajectory. For each pair of $(\mu,\alpha)$ we create a diagram of transitions between states (i.e. 1A,1B,...,3A,3B) in which the thickness of each link corresponds to the probability that that step is taken (Fig. \ref{fig:stepProbability}A). Figure \ref{fig:stepProbability}B shows a graph in which diagrams for different pairs of $(\mu,\alpha)$ have been put together. This graph is very similar to Fig. 3B. This confirms that the most-common trajectories plotted in Fig. 3B are good representatives of the ``typical'' trajectories of successful cells.

\subsection{Adaptation time as function of carrying capacity}

\begin{figure}[h]
   \centering
    \includegraphics[width=0.48\textwidth]{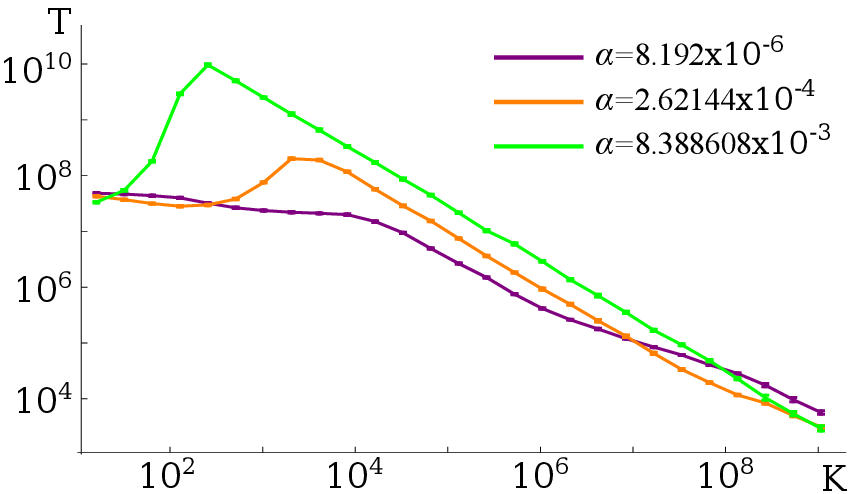}
    \caption{Adaptation time as a function of carrying capacity $K$ for different values of $\alpha$. Remaining parameters are $\mu=10^{-6}, \delta=0.4$ and $d=0.1$}
    \label{fig:timevsK}
\end{figure}

Using the same simulation data as in Fig. 3D but plotting it differently we can explore how the adaptation time $T$ depends on the carrying capacity $K$. Figure \ref{fig:timevsK} shows $T$ for different values of the switching rate $\alpha$. A non-monotonic behaviour can be seen: $T$ is maximal for intermediate $K$ and falls off for both small and large carrying capacities.

\subsection{Evolving the switching rate in fixed increments}

Fig. 5 showed simulation results in which newly created cells could mutate their switching rate $\alpha$ by drawing a new value from a discrete set of allowed switching rates. We observed successful cell trajectories to consist largely of those that evolved switching rates to within the optimal range identified in Fig. 2B. Here we consider the same set of allowed $\alpha$ values but, upon mutation of $\alpha$, the new value is chosen as one of the two closest values to the switching rate of the ancestor cell. This corresponds to $\alpha$ evolving in fixed multiplicative steps, with increase or decrease of $\alpha$ being equally likely. The rest of the algorithm is the same as before.

\begin{figure*}
   \centering
    \includegraphics[width=1.0\textwidth]{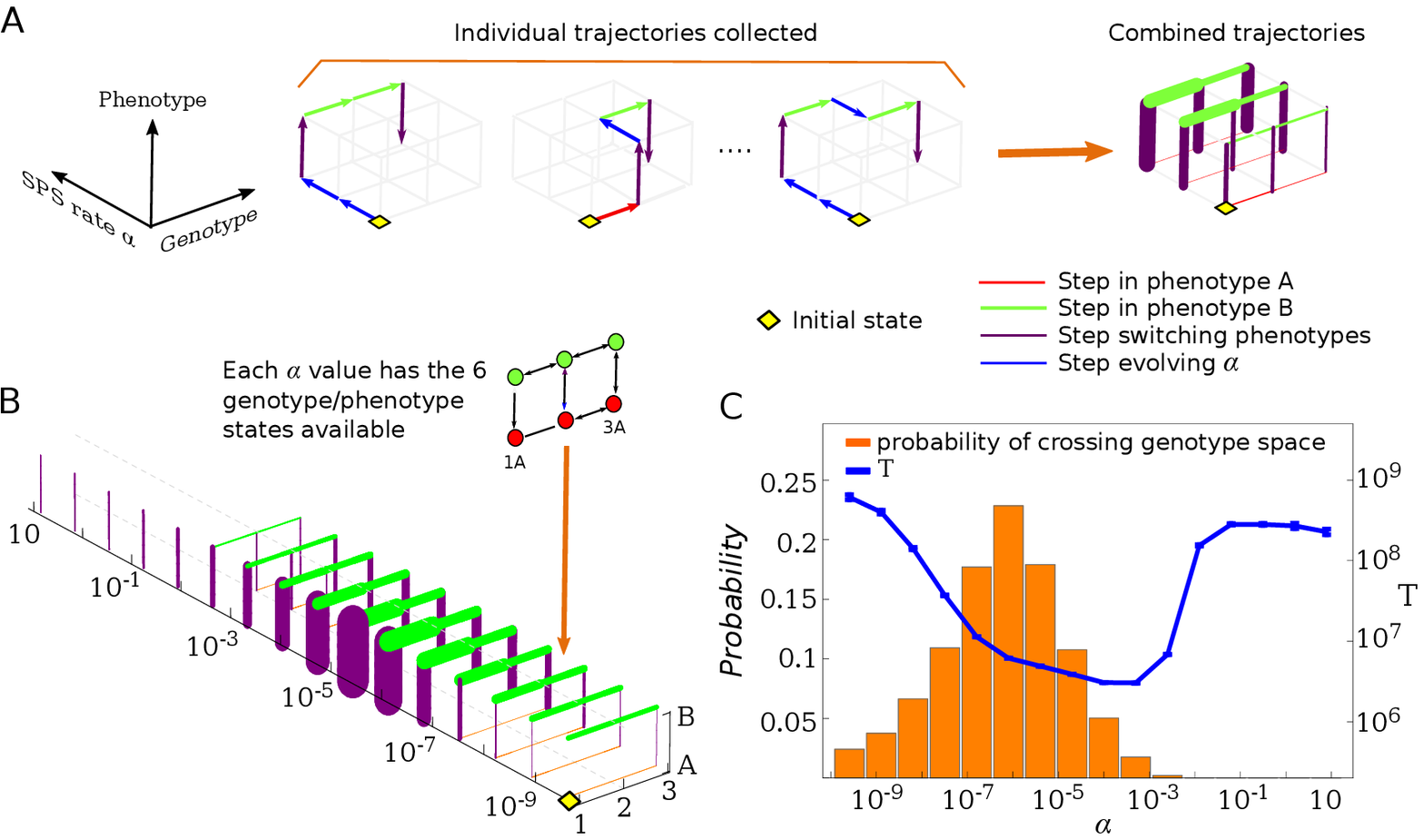}
   \caption{Evolutionary trajectories when $\alpha$ evolves in fixed multiplicative steps. (A) Left: examples of evolutionary trajectories. Upon mutation of $\alpha$ a neighbouring value is selected (with increase or decrease equally likely in the set of allowed values $\alpha=2.56\times10^{-10}\times5^{i}$ for $i\in[0,15]$). (A) Right: trajectories are used to calculate transition probabilities between the states of the system, which are then represented by the thickness of links connecting the states. Red, green and purple links identify steps in phenotype A, phenotype B and between phenotypes respectively. (B) Graph of transition probabilities where line thicknesses are proportional to the probability that a successful trajectory involves that step. The parameters are $K=100, \mu=10^{-5}, \delta=0.4$ and $d=0.1$, the same that were used in Fig. 2B (blue line). The population begins at the wild-type 1A with $\alpha=2.56\times10^{-10}$ and evolves until a cell in state 3A is produced. (C) The probability that genotype space is crossed at a given $\alpha$ in either phenotype A or B. Superimposed on this plot is the adaptation time $T$ as a function of $\alpha$ obtained in a simulation with fixed $\alpha$ (Fig. 2B).}
   \label{fig:EvolvingAlpha2}
\end{figure*}

We again collected trajectories of successful cells in the expanded state space, including states with different $\alpha$, and calculated transition probabilities between any two states. Fig. \ref{fig:EvolvingAlpha2}B shows these probabilities as links of different thickness between the states in state space. We can clearly see some new behaviour here. First, successful trajectories are more likely to feature genotype mutations in phenotype A (more red lines) than in Fig. 5B. Second, the probability of crossing the genotype space at large $\alpha$ (in either phenotype) is much smaller due to the increased number of steps it takes a cell to evolve such large $\alpha$ values. Otherwise, Fig. \ref{fig:EvolvingAlpha2} is similar to Fig. 5. This shows that the details of how mutations of $\alpha$ are implemented do not have a significant effect upon the results of our model.

\end{document}